\documentclass{elsart} 
\usepackage{graphicx}
\begin{document}
\begin{frontmatter}
\title {Development of a Momentum Determined Electron Beam in 
the 1~-~45~GeV Range}
\author[IHEP]{V.A.~Batarin},
\author[FNAL]{J.~Butler},
\author[IHEP]{A.A.~Derevschikov},
\author[IHEP]{Yu.V.~Fomin},
\author[UMN]{V.~Frolov},
\author[IHEP]{V.N.~Grishin},
\author[IHEP]{V.A.~Kachanov},
\author[IHEP]{V.Y.~Khodyrev},
\author[IHEP]{A.S.~Konstantinov},
\author[IHEP]{V.I.~Kravtsov},
\author[UMN]{Y.~Kubota},
\author[IHEP]{V.M.~Leontiev},
\author[IHEP]{V.A.~Maisheev},
\author[IHEP]{Ya.A.~Matulenko},
\author[IHEP]{A.P.~Meschanin},
\author[IHEP]{Y.M.~Melnick},
\author[IHEP]{N.G.~Minaev},
\author[IHEP]{V.V.~Mochalov},
\author[IHEP]{D.A.~Morozov},
\author[IHEP]{L.V.~Nogach},
\author[IHEP]{P.A.~Semenov\thanksref{addr}},
\thanks[addr]{corresponding author, email: semenov@mx.ihep.su}
\author[IHEP]{K.E.~Shestermanov},
\author[IHEP]{L.F.~Soloviev},
\author[IHEP]{V.L.~Solovianov\thanksref{deceased}},
\thanks[deceased]{deceased}
\author[SYR]{S.~Stone},
\author[IHEP]{M.N.~Ukhanov},
\author[IHEP]{A.V.~Uzunian},
\author[IHEP]{A.N.~Vasiliev},
\author[IHEP]{A.E.~Yakutin}
\author[FNAL]{J.~Yarba},
\collab{BTeV electromagnetic calorimeter group}
\date{\today}

\address[FNAL]{Fermilab, Batavia, IL 60510, U.S.A.}
\address[UMN]{University of Minnesota, Minneapolis, MN 55455, U.S.A.}
\address[SYR]{Syracuse University, Syracuse, NY 13244-1130, U.S.A.}
\address[IHEP]{Institute for High Energy Physics, Protvino, Russia}


\begin{abstract}
   A beam line for electrons with energies in the range of 1 to 45 GeV,
low contamination of hadrons and muons and high intensity up to 10$^6$
per accelerator spill at 27 GeV was setup at U70 accelerator in Protvino,
Russia.
A beam tagging system based on drift chambers
with 160~${\mu}$m resolution was able to measure relative electron 
beam momentum precisely. The resolution  $\sigma_p$/p 
was 0.13\% at 
45~GeV where multiple scattering is negligible. This test beam setup
provided a possibility to study properties of lead tungstate 
crystals (PbWO$_4$) for the BTeV experiment at Fermilab.
\end{abstract}
\end{frontmatter}

\section{Introduction}
BTeV is a new experiment being prepared at FNAL, USA \cite{prop}.
It is aimed at challenging the Standard Model explanation of CP-violation,
mixing and rare decays in the {\it b-} and {\it c-}quark systems.
To study final states containing photons, an
electromagnetic calorimeter using lead tungstate (PbWO$_4$)
scintillating crystals will be used.
 The energy resolution of this type of calorimeter
is expected to be better than 1\% for photon (or
electron) energies above 10~GeV. We need to measure the resolution
with a beam able to span a wide range of electron energies and yet having a
low contamination of hadrons and muons. The energy of each beam electron shouldbe known with a
precision significantly less than 1\%. An electron beam
in the energy range of 1 to 45 GeV which satisfies the above requirements
has been commissioned
at the U70 accelerator at Protvino, Russia.

  We determine the energy of each individual electron using a beam tagging
system, since the natural energy spread of the beam is 
$\approx$~3~\% at 27~GeV. 
The tagging system consists of a spectrometer magnet
and four drift chamber stations. In order to decrease multiple
scattering, the beam was transported in vacuum. This system was able to measure
the beam particle momentum with a
precision of 0.13\% at 45~GeV, where the contribution of multiple
scattering is negligible. At 1~GeV a precision of momentum measurements was
2\%, where multiple scattering dominated. 

   It is worth noting that, we measure the absolute value of the beam momentum
to 1-2\% accuracy, not as well as the above resolution.
It is due to the accuracy of the spectrometer magnetic field measurements.
For our  studies of the electromagnetic calorimeter prototype we do 
not need to know 
the absolute value of the beam energy with a very high precision since the
energy resolution does not strongly depend on energy.     

  The high intensity of the electron beam (up to $1\times10^6$ per accelerator
spill at 27~GeV) and low background provide a good environment to study
crystal radiation hardness. The same beam channel was able to provide a
high intensity (up to $10^7$ per accelerator spill) 40~GeV pion beam.

\section{General beam setup}
   
\begin{figure}
\centering
\includegraphics[width=0.95\textwidth]{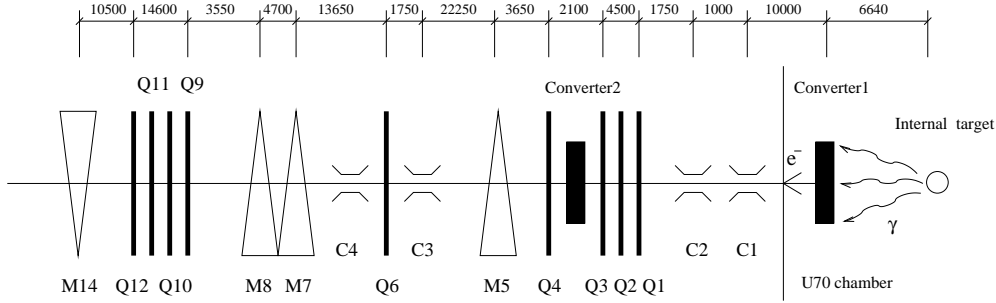}
\caption{Beam optics diagram. Q - quadrupole lenses, M - dipole magnets,
C - collimators. Distances are in mm.}
\label{fig:beam_optics}
\end{figure}

      A method to obtain an electron beam from U70 proton beam is
based on decays of neutral mesons (mainly $\pi^o$-mesons)
from the proton beam interactions on an internal carbon 
film target \cite{e_beam}.  
Photons from $\pi^o$-decays are converted  to electrons in an 
additional target (Converter1) made of $\approx$3 mm 
(0.7X$_0$) thickness Pb plate placed 6.6~m from the internal target. 
(see Fig.\ref{fig:beam_optics}). Charged hadrons from the 
internal target are swept away  from the $\gamma$-quanta path 
by the U70 accelerator magnetic field and don't enter the beam line.
A small number of hadrons  in the electron beam are produced by 
neutron interactions with  Converter1. Electrons 
from Converter1 are guided by the accelerator magnetic field to 
the beam line entrance. The output electron beam momentum 
is defined by the radial position of the internal target. The Converter1
placement in the accelerator chamber allows to obtain more intense beams
($10^6$ electrons per $10^{12}$ protons) 
but reduces the available electron beam momentum range to 25-45~GeV.  
 

    A further decreasing of initial momenta of electrons was obtained 
using an oriented silicon crystal \cite{nim216} (Converter2)
which was placed before the analyzing magnet M5. 
The electron beam after Converter2 has a wide momentum spectrum due to
Bremsstrahlung radiation. The fields of the magnetic elements are 
setup proportionally to a beam momentum.  Finally we have electron beam with 
any required energy from ~1 GeV up to ~45 GeV. 
It usually took us about 15 minutes to tune the beam line elements for 
given energy.


\section{Beam tagging system}
\begin{figure}[b]
\centering
\includegraphics[width=0.95\textwidth]{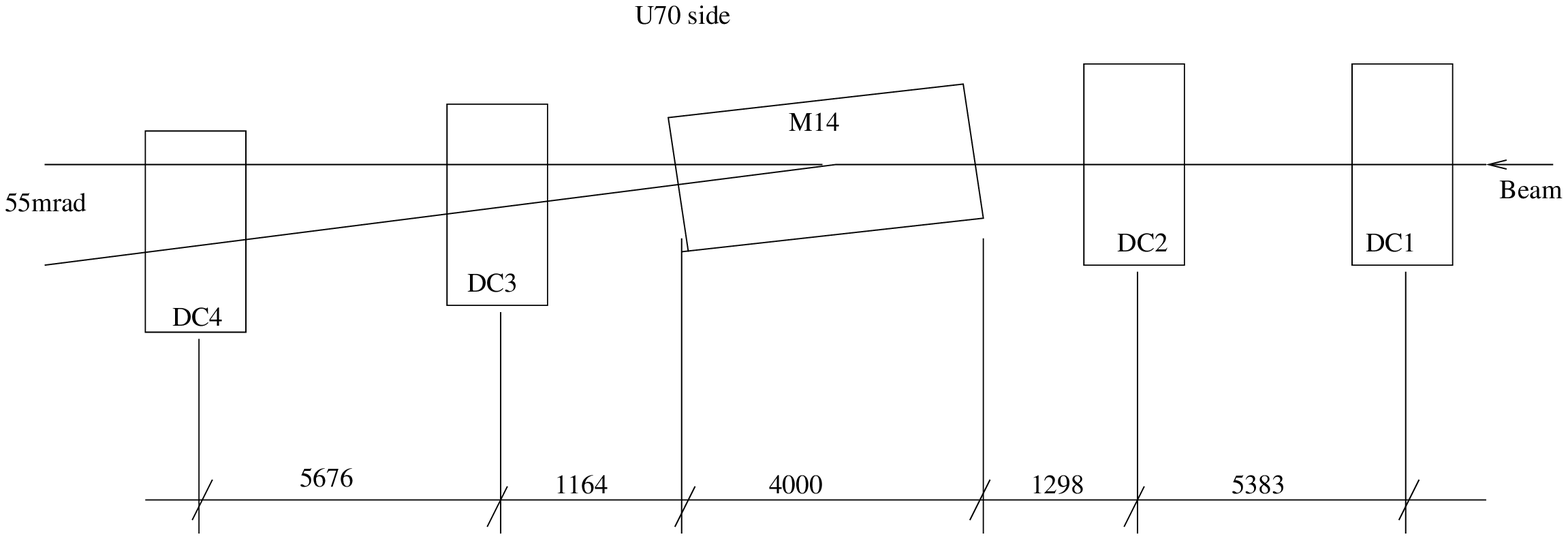}
\caption{The beam momentum tagging system. DC indicates a set 
of drift chambers,
while M14 is a dipole bending magnet. (All distances are in mm.)}
\label{fig:dc_setup}
\end{figure}

   The beam momentum spread was $\Delta$p/p $\approx$~3~\% at 27~GeV 
and was too large to study the energy resolution of PbWO$_4$ crystals. So 
we decided to install momentum tagging system to measure the momentum of 
each electron. It consisted of four drift chamber
stations (DC) and a 4 meter long spectrometer magnet denoted M14
(see Fig.~\ref{fig:dc_setup}).
The magnet deflected the beam in the horizontal plane by 55~mrad.
The magnet current was adjustable in order to provide the same bending
angle for all energies of the electron beam.

The {\it X} and {\it Y} positions of the charged particle were
measured in each DC station with a pair of drift chambers in each view, 
which shared the same gas volume.
Each chamber has a $20\times 20~$cm$^2$ sensitive area. 
The third station, DC3, has only a pair of chambers measuring the {\it x} 
coordinate.
All together four DC stations consisted of fourteen drift chambers. 

\begin{figure}
\centering
\includegraphics[width=0.7\textwidth]{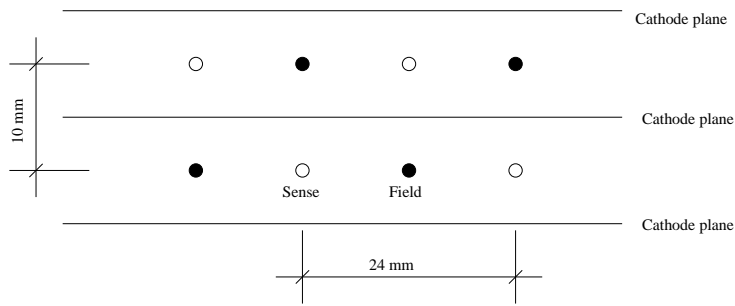}
\caption{The structure of each drift chamber doublet for a single view.}
\label{fig:dc_struct}
\end{figure}

The internal structure of a pair of drift chambers
is shown in Fig.~\ref{fig:dc_struct}.
\begin{figure}
\centering
\includegraphics[width=0.9\textwidth, height=0.7\textwidth]{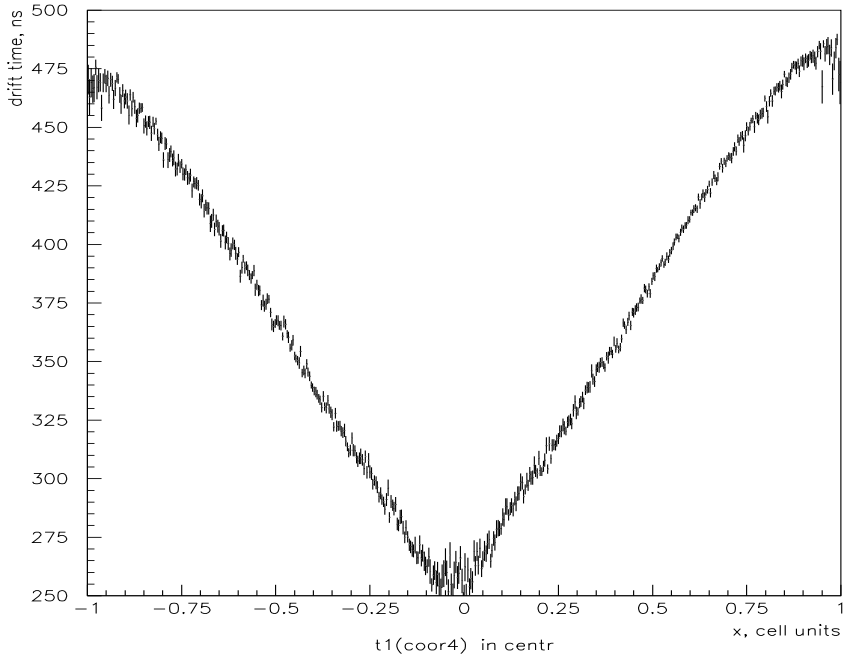}
\caption{Drift time versus track position dependence.}
\label{fig:t_r_func}
\end{figure}
Each drift cell is formed by a signal (sense) wire in the center, field
wires at the edges of the cell  and cathode planes perpendicular to the beam
direction.  The signal wires are separated by 2.4~cm  corresponding 
to a maximum
drift distance of 1.2~cm. The signal wires of the second chamber were
shifted by 1.2~cm to resolve the left-right ambiguity. The distance between
signal wires and cathode planes is 5~mm. The signal wires were made of
gold plated tungsten (20~$\mu m$ diameter) and field wires,
beryllium-copper alloy (100~$\mu m$ diameter). The cathode planes were made of
graphite coated (5~$\mu m$) mylar (total thickness 25~$\mu m$).
To complete the electrostatic shielding of the modules, aluminized mylar
(25~$\mu m$ mylar and less then 1~$\mu m$ Al) was placed
on the sides of the box.

 Each chamber required two high voltages: the cathode voltage (HVC), and
the voltage on the field wires (HVF) to make
a uniform electric field along the cell.  During
the experiment all chambers were operated with HVC= -1.8 kV and
HVF= -2.9 kV.

\begin{figure}
\centering
\includegraphics[width=0.7\textwidth]{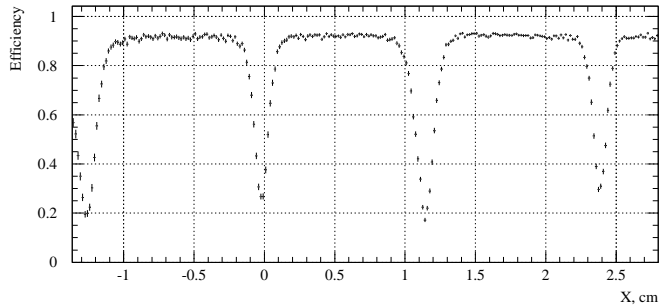}
\caption {A cell efficiency for the x-plane of the drift chamber DC2.
       X=0 corresponds to the sense wire position in the first subplane.
       Minima are caused by and correspond to the field wire positions
       both in the first and shifted subplanes.}
\label{fig:dceff}
\end{figure}

 A mixture of argon (70\%) and isobutane (30\%) gas was used.
The operating voltages gave a field gradient of 1~kV/cm in the
main part of the drift space, which is enough to provide for a saturated
drift velocity.

 The design of the DC internal structure and operating voltages were the 
result of a compromise
between predictions of a GARFIELD simulation for a constant drift velocity
within the drift cell, track detection efficiency, and our
requirement of minimum materials. 

Assuming that the drift velocity is constant over the entire cell, 
we achieved the required track precision. 
The actual measured time distribution function for one of the drift chambers, 
$t(x)$, is shown in Fig.~\ref{fig:t_r_func} where {\it x} was obtained 
from a track fit over the three other DC stations.
It shows some non-uniformity, so the precision of the track
reconstruction can be improved further by fitting the drift velocity as a
function of {\it x}. The DC efficiency is shown in Fig.~\ref{fig:dceff}. We
reach the $\sim$90\% level over most of the cell. 
The drift time was measured by LeCroy~3377 TDC with a 1~ns/count accuracy.

\section{Contamination of hadrons and muons in the electron beam}

      The fraction of electrons in the beam for various momenta
were estimated in the following way.
At the end of the beam line, after DC4 (see Fig.\ref{fig:dc_setup}), we place
a box containing a 5$\times$5 matrix of lead tungstate
scintillating crystals. Using information from the
drift chambers, beam particles  which hit within a $3\times 3$ region in the
center of the  crystal matrix have
been selected. The crystals are square, 27 mm on each of the lateral side and
220 mm in length. 
For these events, the energy deposit in the entire
calorimeter was measured. 

\begin{figure}[th]
\centering
\includegraphics[width=0.9\textwidth]{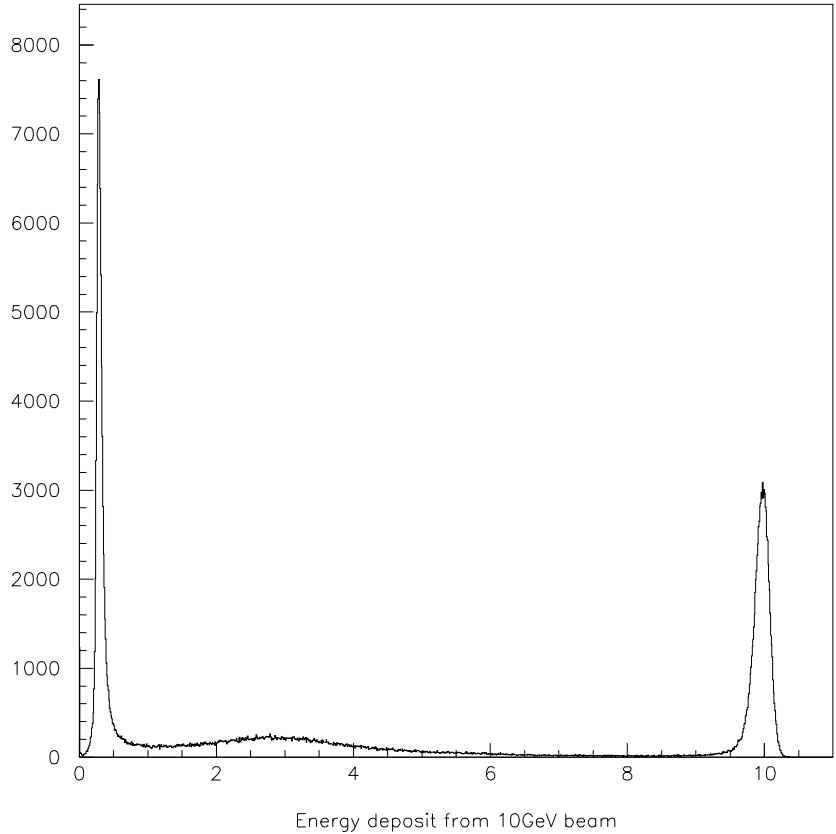}
\caption{Energy spectrum of the 10 GeV electron beam. This is the worst
case in terms of a purity of the obtained electron beam.}
\label{fig:mom10}
\end{figure}

Fig \ref{fig:mom10} shows the distribution of these measured energies
for 10 GeV electron beam data. This is the worst case in terms of 
electron beam purity. We see clean muon and electron peaks at
about $\approx 0.3~$GeV and $\approx 10~$GeV, respectively and the energy
deposit from  hadronic showers between them. The relatively large muon fraction
was useful to monitor the stability of the
calorimeter prototype properties.

The events with the energy deposit
greater than 0.9 of MEAN value of the peak corresponding to the
beam energy have been assigned to electrons. The rest of the events are
background particles, muons and hadrons. The measured
fractions of electrons in the beam for each beam energy are shown
in Table~\ref{tab:frac}. All fractions are determined to an absolute 
accuracy of 1\%.

\begin{table}[hbt]
\caption{Fraction of electrons in the beam at various energies}
\label{tab:frac}
\begin{center}
\begin{tabular}{cccccc}
\hline\hline
 1~GeV & 2~GeV & 5~GeV & 10~GeV & 27~GeV & 45~GeV \\
 82\% & 77\%   & 50\%  & 34\%   & 77\%   & 91\% \\
\hline
\end{tabular}
\end{center}
\end{table}

Note that by selecting events using drift chamber information, we rejected
all charged particles which have momenta outside of the tagging station
fiducial region. So, the obtained result does not reflect the total 
contamination
of the hadrons and muons in the electron beam, 
which is estimated to be 25\% more for 27 GeV.

\section{Precision of the beam momentum measurement}

In this section we estimate the accuracy of the beam momentum determination
with our spectrometer. The design of the spectrometer was optimized
for precision charged-particle tracking and momentum determination,
as well as for precise position  measurements of high
energy electrons in the lead tungstate $5\times 5$ electromagnetic
calorimeter prototype.


  The momentum resolution $\sigma_p/p$ of the spectrometer
equals the accuracy $\sigma_\theta/\theta$, where $\theta$ is the 
deflection angle in the analyzing a magnet. This angle is calculated from 
the linear combination of the 4 position measurements:
\begin{equation}
\theta \approx \sum_{i=1}^{4}{b_i \cdot x_i + c_{survey}}
\end{equation}
where $b_i$ are the coefficients which depend on the distance  of the 
chambers from the analyzing magnet center, $x_i$'s are transverse 
coordinates of the beam particle 
measured in the drift chambers, and $c_{survey}$ is a survey constant.

Two factors contribute to $\sigma_\theta$, the uncertainty in the $\theta$
measurements.
The first one is a drift chamber resolution. The second is a multiple
Coulomb scattering on materials in the beam line. The main contribution 
of Coulomb scattering to the $\sigma_\theta$ is given by materials of
the DC2 and DC3 stations since they are near the 
magnet (see Fig.\ref{fig:dc_setup}).

If we assume that all the drift chambers have the same position
resolution ${\sigma_x}$, then
\begin{equation}
\sigma^2_{\theta} = \sigma^2_x \cdot \sum_{i=1}^{4}{b_i^2 +
\sigma^2_{\theta_{Coulomb}}}
\label{eq:sigma2t_full}
\end{equation}

\begin{figure}[bh]
\centering
\includegraphics[width=0.95\textwidth]{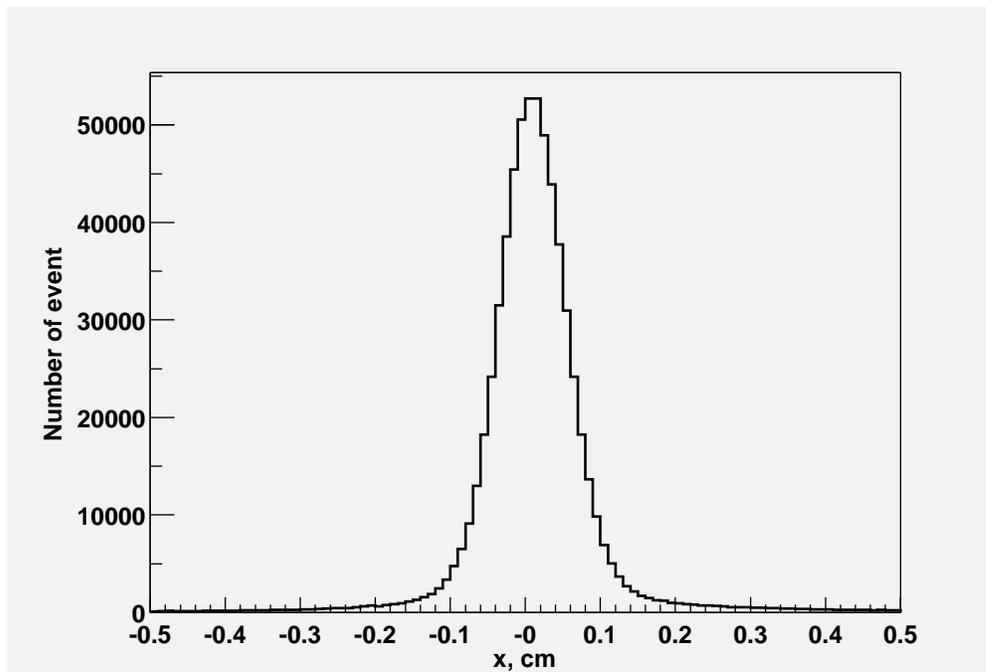}
\caption{Distribution of the lateral position difference in the center of 
the analyzing magnet for trajectories reconstructed by the upstream 
(DC1 and DC2) and downstream (DC3 and DC4) chambers.}
\label{fig:x45}
\end{figure}

To define $\sigma^2_\theta$ at different energies one needs to know
the drift chamber position resolution and the mean multiple scattering
angle.
These values can be determined experimentally. We extrapolated the beam 
trajectories reconstructed by the upstream drift chambers (DC1 and DC2) 
and downstream drift chambers (DC3 and DC4) to the center of the 
analyzing magnet. Their lateral position (distance to the nominal beam 
position) should agree. We defined their difference as $D_x$, which can
be calculated from another linear combination of the four 
{\it x} measurements of each beam 
particle and expressed in equation \ref{eq:D_x}. The distribution of $D_x$ 
for a number of beam particles is shown in Fig.\ref{fig:x45}.

\begin{equation}
            D_x= \sum_{i=1}^{4}{a_i \cdot x_i}
\label{eq:D_x}
\end{equation}
where $a_i$ are the coefficients which depend on the distance of the chambers
from the analyzing magnet
 and $x_i$ are transverse position of the beam particle measured in the drift
chambers. We adjusted the maximum drift time so that the width of the 
$D_x$ distribution is the narrowest. The quantity of the maximum drift time
gives us the drift velocity.

The width of $D_x$ distribution is related to 
the resolution of the chamber stations, $\sigma_x$, and Coulomb 
multiple scattering effects by
\begin{equation}
\sigma^2_{D_x}=\sigma^2_x \sum_{i=1}^{4}{a^2_i} + \sigma^2_{\theta_{Coulomb}}
\cdot Z^2_{eff}
\label{eq:sigmaDx}
\end{equation}

\begin{equation}
\sigma^2_{\theta_{Coulomb}}\cdot Z^2_{eff} =
\left(\frac{E_0}{E}\right)^2(t_2\cdot z^2_2 +
t_3\cdot z^2_3) = \left(\frac{E_0}{E}\right)^2\cdot t \cdot \frac{t_2\cdot z^2_2
+
t_3 \cdot z^2_3}{t_2 + t_3},
\end{equation}
where $t_2$ and $t_3$ are the relative material thickness  
around the drift chambers DC2 and DC3 in radiation length units ($x_i/X_0$),
$t$ is the full thickness, $t = t_2+t_3$, $z_i$ is the distance from the 
drift chamber to the center of the magnet, {\it E} is an electron energy, 
$E_0$ is equal to 13.6~MeV.
$Z_{eff}$ is the effective distance from the center of the magnet to
the scattering center given by
\begin{equation}
\frac{t_2\cdot z^2_2 +t_3 \cdot z^2_3}{t_2+t_3} = Z^2_{eff} \approx
\frac{z^2_2 +z^2_3}{2} \approx (3.3~m)^2,
\end{equation}
since $t_2 \approx t_3$.
The equation (\ref{eq:sigmaDx}) can be rewritten in the following way
\begin{equation}
\sigma^2_{D_x} = S + \frac{C}{E^2},
\end{equation}
emphasizing the difference in the beam energy dependence of the two terms.
We see that $\sigma^2_{D_x}$ is a linear function of $1/E^2$.
We have the measurements of $\sigma^2_{D_x}$ for a set of energies.
Fitting these data by a straight line we have found that
\begin{displaymath}
\sigma^2_{D_x} = 0.219 + \frac{14.22}{E^2}
\end{displaymath}
Therefore the position resolution of the DC station
assuming that all the chambers have the same resolution is as
follows:
\begin{equation}
\sigma^2_x = \frac{S}{\sum^4_{i=1}{a^2_i}}
\label{eq:sigma2x}
\end{equation}

\begin{table}
\caption{MC results and measured momentum detection precision. $\sigma^{*}_p$
represents FWHM/2.35.}
\begin{tabular}{|c|c|c|c|c|}
\hline
 &\multicolumn{2}{c|}{Monte Carlo simulation}&\multicolumn{2}{c|}{Measured
values}\\
\cline{2-5}
 & {\small No errors in DC} & {\small $\sigma_x$=0.16 mm} & resolution & {\small
Beam momentum spread} \\

\hline
Momentum, GeV/c & $\sigma_p/p$, \% & $\sigma_p/p$, \% & $\sigma_p/p$, \% &
$\sigma^{*}_p/p$, \%\\
\hline
  1.0   &      2.38$\pm$0.04    &       2.38$\pm$0.04   & 2.05 & 4.3 \\
\hline
  2.0   &      1.14$\pm$0.04   &        1.14$\pm$0.04   & 1.03 & 5.5 \\
\hline

   5.0 &    0.52$\pm$0.01  &       0.53$\pm$0.01 & 0.43 & 5.6  \\
\hline
  10.0 &     0.25$\pm$0.01      &        0.28$\pm$0.01  & 0.24 & 3.8  \\
\hline
  26.7 &      0.096$\pm$0.002    &        0.149$\pm$0.002  & 0.15 & 1.2 \\
\hline
  45.0 &      0.067$\pm$0.006    &        0.126$\pm$0.006  & 0.13 & 1.0 \\
\hline
\end{tabular}
\label{tab:beam_mom}
\end{table}

\begin{equation}
\sigma^2_{\theta_{Coulomb}} = \frac{C}{Z^2_{eff}} \cdot \frac{1}{E^2}
\label{eq:sigma2t}
\end{equation}
From (\ref{eq:sigma2x}) we found $\sigma_x$ = 0.16~mm,
which corresponds to a single chamber resolution to be 160*$\sqrt{2}$.
When we put (\ref{eq:sigma2x}) and (\ref{eq:sigma2t}) in equation
(\ref{eq:sigma2t_full}), we obtain
\begin{equation}
\sigma^2_\theta = \frac{S}{\sum^4_{i=1}{a^2_i}} \cdot \sum^4_{i=1}{b^2_i} +
\frac{C}{Z^2_{eff}} \cdot \frac{1}{E^2}
\label{eq:sigma2th}
\end{equation}

The results of the calculations using (\ref{eq:sigma2th}) are given in column
4 of the Table \ref{tab:beam_mom} as the measured values for a  beam 
momentum resolution. The last column of the same table
shows the measured by our spectrometer beam momentum spread ($\sigma^{*}_p$/p) 
in \%, where $\sigma^{*}_p$ represents FWHM/2.35 of the main peak 
since the momentum distribution is non-Gaussian.

     The geometry of this beam line was reproduced in a GEANT
Monte Carlo simulation taking into account the real materials distribution.
The Moliere theory, corrected for finite angle
scattering, was used to calculate the effect of multiple Coulomb scattering
on the charged particle trajectory. The position resolution of each drift
chamber station was taken as a Gaussian distribution 
with $\sigma_x=\sigma_y=0.16$~mm.
The momentum resolution was reconstructed using the same cuts as the real
data. The results are presented in Table \ref{tab:beam_mom}.
Where, $\sigma_p/p$ is the relative error of the electron momentum.
$\sigma_x$ and $\sigma_p$ are the r.m.s. of the Gaussian fit of
the corresponding distribution.

\begin{figure}
\centering
\includegraphics[width=0.95\textwidth]{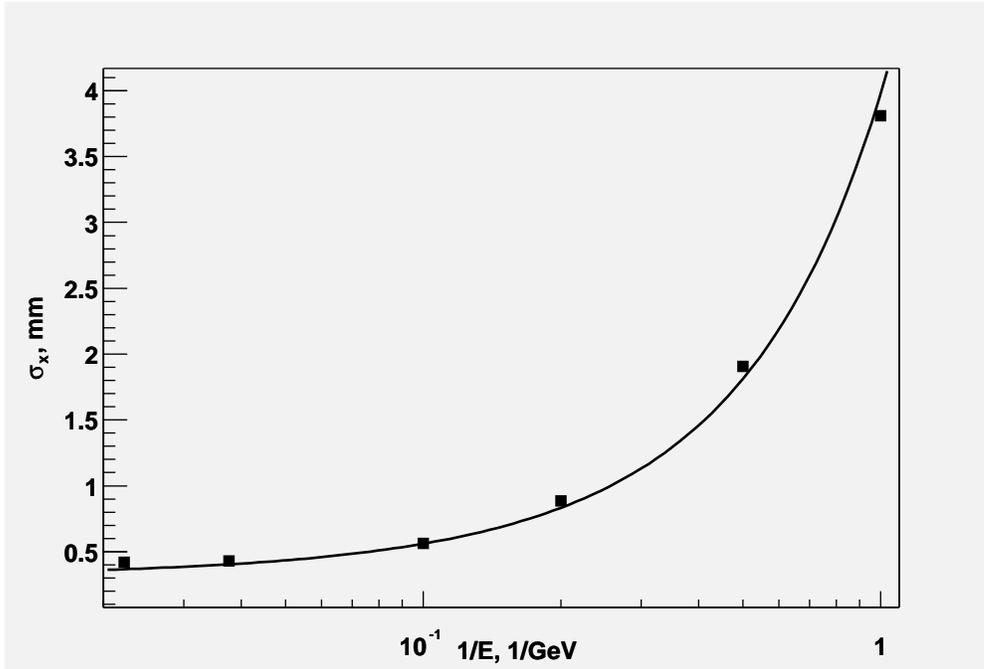}
\caption{Lateral distance distribution RMS dependence on energy.
Solid line corresponds to MC-simulation results,
filled squares real measurements.}
\label{fig:sigmap}
\end{figure}

The dependence of $\sigma_{D_x}$ on energy was also simulated by GEANT. 
The results
are shown in the Fig. \ref{fig:sigmap}. Solid line represents
Monte Carlo simulation, filled squares real measurements. Estimated errors
of the experimental data are inside the markers.
There is a good agreement between the simulation data and the 
real measurements which are shown in 
Table \ref{tab:beam_mom} and Fig.\ref{fig:sigmap}. It proves the validity of
our  drift chamber resolution determination described above.

\section{Conclusion}
The experience of long-term operation of a modified beam line
for electrons with energies in the range of 1 to 45 GeV
described in this paper shows that this technique significantly
expands the possibilities to study precise energy and coordinate
resolutions of scintillating crystals. Independent of the momentum spread
of the electron beam at the level of 1 to 5\% at energies from 45
down to 1 GeV, the momentum tagging station gives a beam
momentum resolution from 0.13 to 2\% in the same energy range.
The precision is limited by drift chamber
spatial resolution and multiple Coulomb scattering on a material
in the beam line.
GEANT Monte Carlo simulations of the resolution agree very well with the
experimental results.

   An opportunity to switch the beam line from electrons to high
energy pions with high intensities (up to 10$^6$ e$^-$
and 10$^7 \pi^-$ per spill) allows the study radiation
hardness properties of scintillating crystals both for electrons
and hadrons with moderate dose rates of up to 100 rad/hour. 
These dose rates are similar to the ones
that will be in BTeV at Fermilab and in CMS at CERN.
Integrated doses up to several krad can be accumulated in this
setup in a relatively short time.

\section{Acknowledgments}
    We would like to thank the IHEP management for providing us
a beam line and accelerator time for our testbeam studies.
Special thanks to Fermilab for providing
equipment for data acquisition. This work was partially supported by the U. S.
National Science Foundation and the Department of Energy.

\end{document}